\documentclass{ws-procs9x6-cpt16}
\begin{document}

\newcommand{\refeq}[1]{(\ref{#1})}
\def\etal {{\it et al.}}
%any other macros go here 
\newcommand{\vect}[1]{\boldsymbol{#1}}

\title{Limit on Lorentz-Invariance- and CPT-Violating Neutron Spin Interactions Using a $^3$He-$^{129}$Xe Comagnetometer}

%%%%%%%%%%%Author list
\author{F. Allmendinger\footnote{Corresponding author: allmendinger@physi.uni-heidelberg.de}, U. Schmidt}
\address{Physikalisches Institut, Ruprecht-Karls-Universit\"{a}t, 69120 
Heidelberg, Germany}

\author{W. Heil, S. Karpuk, and Yu. Sobolev}
\address{Institut f\"{u}r Physik, Johannes Gutenberg-Universit\"{a}t, 
55099 Mainz, Germany}

\begin{abstract}
We performed a search for a Lorentz-invariance- and CPT-violating coupling of the $^3$He and $^{129}$Xe nuclear spins to posited background fields. Our experimental approach is to measure the free precession of nuclear spin polarized $^3$He and $^{129}$Xe atoms using SQUID detectors. As the laboratory reference frame rotates with respect to distant stars, we look for a sidereal modulation of the Larmor frequencies of the co-located spin samples. As a result we obtain an upper limit on the equatorial component of the background field $\tilde{b}^n_{\bot}< 8.4 \cdot 10^{-34}$ GeV (68\% C.L.). Furthermore, this technique was modified to search for an electric dipole moment (EDM) of $^{129}$Xe.
\end{abstract}

\bodymatter
%%%%%%%%%%%%%%%%%%%%%%%%%%%
\section{Introduction and Experimental Setup}
Nuclear spin clocks, based on the detection of free spin precession of gaseous nuclear polarized $^{3}$He and $^{129}$Xe atoms with LT$_C$ SQUIDs as low-noise magnetic flux detectors are used as ultra-sensitive probe for nonmagnetic spin interactions, since the magnetic interaction (Zeeman term) drops out in the case of co-located spin samples (comagnetometry). Measurements of uninterrupted precession of one day can be achieved at the present stage of investigation due to long spin-coherence times. The principle of measurement is to search for sidereal variations of the precession frequency of co-located spin species while the Earth and hence the laboratory reference frame rotates with respect to distant stars. \\
In the context of the Standard-Model Extension (SME)\cite{Kostelecky,Colladay,Kostelecky2}, couplings of the neutron or proton spin $\vect{\sigma^{n, p}}$ to relic background fields $\vect{\tilde {b}^{n,p}}$ are discussed. The background fields have distinct directions in space and correspond to preferred spin directions. These couplings with the potential $V=\vect{\tilde {b}^{n,p}}\cdot \vect{\sigma^{n, p}} $ are purely non-magnetic, but change the energy levels of spins in a magnetic field, which can be detected by changes in the Larmor frequency of precessing spins.\\
To give a short overview of the setup (details are given in Ref.~\cite{Allmendinger}): The two polarized gas species (and N$_2$ as a buffer gas) are filled into a low-relaxation spherical glass cell with radius $R=5$~cm. Typically, the optimum conditions in terms of long transverse relaxation times $T_2^*$ and high Signal-to-Noise Ratio are met at a gas mixture with pressures of $p_{He}=3$~mbar, $p_{Xe}=5$~mbar, $p_{N2}=25$~mbar. The cell is positioned in a homogeneous static magnetic field of about 400~nT that is generated by Helmholtz coils mounted inside the strongly magnetically shielded room \textit{BMSR-2} at the \textit{Physikalisch-Technische Bundesanstalt} in Berlin.
At that field strength, the Larmor frequencies of $^{3}$He and $^{129}$Xe are about $\omega _{He}\approx 2\pi \cdot 13$~Hz and $\omega _{Xe}\approx 2\pi \cdot 4.7$~Hz, respectively. To measure these precession frequencies very precisely, low-noise low-temperature DC-SQUID gradiometers are used as magnetic flux detectors. Due to the very low field gradients in the order of pT/cm at the location of the cell, the transverse relaxation times reached $T_2^*=8.5$~h for $^{129}$Xe and up to $T_2^*=100$~h for $^{3}$He~\cite{Allmendinger}. The measured signal amplitudes at the beginning of the measurement were up to $A_\text{He}=20$~pT and $A_\text{Xe}=8$~pT for $^3$He and $^{129}$Xe, respectively. The noise level (combination of four gradiometers) was $\rho=3$~fT/$\sqrt{\text{Hz}}$. Due to the long spin-coherence time and the high initial Signal-to-Noise Ratio, the spin precession could be monitored for more than one day, which improves the sensitivity remarkably.
%%%%%%%%%%%%%%%%%%%%%%%%%%%%%%%%%%%%%%%%%%%%
\section{Data Evaluation and Results}
To be sensitive to tiny nonmagnetic interactions, one has to consider the weighted difference of the respective Larmor frequencies of the co-located spin samples, or the corresponding time integral, the weighted phase difference, which are defined by
\begin{equation}
\label{eqn:weightedpd}
\Delta \omega =\omega_{\rm He} -\frac{\gamma_{\rm He} }{\gamma_{\rm Xe} }\omega _{\rm Xe}~~\text{and}~~\Delta \Phi =\Phi_{\rm He} -\frac{\gamma_{\rm He} }{\gamma _{\rm Xe} 
}\Phi _{\rm Xe}~~.
\end{equation}
In doing so, magnetic field fluctuations are canceled, i.e. in principle $\Delta \omega =0$ and $\Delta \Phi=$const. if there are no further interactions. However, on a closer look, $\Delta \Phi $ is not constant in time, as 
higher order effects have to be take into account. These can be parameterized by
\begin{eqnarray}
 \Delta \Phi (t)&=&c_0 +c_1t+E_{\rm He} e^{-t/T_{\rm 2,He}^\ast 
}+E_{\rm Xe} e^{-t/T_{\rm 2,Xe}^\ast } +F_{\rm He} e^{-2t/T_{\rm 2,He}^\ast }+F_{\rm Xe} e^{-2t/T_{\rm 2,Xe}^\ast }~~.
\label{eq2}
\end{eqnarray}
The linear contribution stems from Earth's rotation (i.e. the rotation of the SQUID detectors with respect to the precessing spins) and from chemical shift (diamagnetic shielding: the electron shells shield the nuclei against the external magnetic guiding field). The four exponential terms account for the Ramsey-Bloch-Siegert shift \cite{Bloch,Ramsey}. These effects are discussed in Ref.~\cite{Allmendinger}.
Finally, the function in Eq. (\ref{eq2}) together with the appropriate parameterization of the Lorentz-invariance-violating effect - in this case a sidereal modulation $\propto \tilde{b}\cdot\sin(\Omega_S\cdot t+\varphi_0)$ - is fitted to the combined weighted phase difference data of all measurement runs (7 in total). The resulting estimate on sidereal modulation is compatible with zero within the correlated and uncorrelated uncertainties and can be expressed as an upper limit on the magnitude of the hypothetical background field:
\begin{eqnarray}
\label{eqn:results}
\tilde{b}^{\text{n}}_{\bot}&<&6.7 \cdot 10^{-34}\text{ GeV}  \text{ (68\% C.L.)~~.}
\end{eqnarray}

In Ref. \cite{Stadnik}, Y. Stadnik and V. Flambaum showed that the $^3$He-$^{129}$Xe comagnetometer is also sensitive to the proton interaction parameters of the SME \cite{Stadnik}. Based on our measurements the following values were derived:
\begin{eqnarray}
\label{eqn:protonresults}
\nonumber \tilde{b}^{\text{n}}_X+0.74\cdot\tilde{b}^{\text{p}}_X&=&(7.1\pm8.2)\cdot 10^{-34}\text{ GeV}\\
\tilde{b}^{\text{n}}_X+0.74\cdot\tilde{b}^{\text{p}}_Y&=& (5.0\pm 10.8) \cdot 10^{-34}\text{ GeV}~~~.
\end{eqnarray}
The corresponding upper limit of the equatorial component $\tilde{b}^{\text{p}}_{\bot}$ of the background tensor field interacting with the spin of the bound proton is
\begin{eqnarray}
\label{eqn:protonresults2}
\tilde{b}^{\text{p}}_{\bot}&<1.6 \cdot 10^{-33}\text{ GeV}  &\text{(68\% Confidence Level)}~.
\end{eqnarray}
%%%%%%%%%%%%%%%%%%%%%
\section{Conclusion and Outlook}
Freely precessing gaseous, nuclear polarized $^{3}$He and $^{129}$Xe samples can be used as ultra-sensitive probe for nonmagnetic spin interactions, since the magnetic interaction (Zeeman term) drops out in the case of co-located spin samples. With a similar setup, upper limits on interactions mediated by axion-like particles were obtained~\cite{Tullney}.\\
The next step is to apply this method to search for a CP-violating permanent electric dipole moment (EDM) of $^{129}$Xe: A permanent EDM $\vect{d}$ of a fundamental or composite particle must be aligned parallel to the spin, as the spin is the only available vector for an eigenstate of the isolated particle. Thus, for a magnetic guiding field aligned along the z-direction, the Hamiltonian has the form $H=-\mu\cdot B_0-d \cdot E_z$, with the corresponding frequency shift $\delta\omega_\text{EDM}=\frac{2}{\hbar} d \cdot E_z$. By varying the z-component of the electric field $E_z$, the frequency shift is modulated correspondingly. As mentioned before, the principle of comagnetometry is applied to become insensitive to drifts of the magnetic guiding field. Subsequently, a non-zero EDM will manifest in a modulation of the weighted phase difference, and the corresponding value $d$ can be extracted.\\
The experimental setup has been changed to enable the measurement of the $^{129}$Xe EDM: The measurement cell has a cylindrical shape. The end planes are made of silicon and the lateral surface is composed of the low-relaxation GE-180 glass. The distance between the electrodes is 5~cm and the maximum voltage that can be applied is 12~kV. A small amount of SF$_6$ (a few mbar) is added to the gas mixture to suppress leakage currents. It is highly beneficial to maximize the coherent measurement time $T$, as the uncertainty in frequency determination (and thus the error on the EDM) decreases as $T^{-3/2}$ for white noise. One important mechanism that reduces the spin-coherence time is caused by magnetic field gradients across the measurement cell. Consequently, additional gradient coils have been integrated into the setup to compensate residual magnetic field gradients of the mu-metal shielding.

\end{document}